\documentclass[prl,aps,floats,twocolumn]{revtex4}

\usepackage[dvips]{graphicx}
\usepackage{amssymb}
\usepackage{amsmath}

\usepackage{color}

\begin{document}

\title{Can binary mergers produce maximally spinning black holes?}

\author{Michael Kesden}

\affiliation{Canadian Institute for Theoretical Astrophysics
  (CITA), University of Toronto, 60 St.~George Street, Toronto,
  Ontario, M5S 3H8, Canada}

\date{July 2008}
                            
\begin{abstract}
  Gravitational waves carry away both energy and angular momentum as
  binary black holes inspiral and merge.  The relative efficiency with
  which they are radiated determines whether the final black hole of
  mass $M_f$ and spin $S_f$ saturates the Kerr limit ($\chi_f \equiv
  S_f/M_f^2 \leq 1$). Extrapolating from the test-particle limit, we
  propose expressions for $S_f$ and $M_f$ for mergers with initial
  spins aligned or anti-aligned with the orbital angular momentum.  We
  predict the the final spin at plunge for equal-mass non-spinning
  binaries to better than 1\%, and that equal-mass maximally spinning
  aligned mergers lead to nearly maximally spinning final black holes
  ($\chi_f \simeq 0.9988$).  We also find black holes can always be
  spun up by aligned mergers provided the mass ratio is small enough.
\end{abstract}
\maketitle


Recent breakthroughs in numerical relativity have led to successful
simulations of the inspiral, merger, and ringdown of binary black
holes (BBHs) \cite{Pretorius:2005gq,Baker:2005vv,Campanelli:2005dd}.
Despite this progress, approximations underlying these simulations
limit them to mergers with large but submaximal initial spins ($\chi_i
\lesssim 0.9$) and order-unity mass ratios ($q \equiv m_2/m_1 \geq
1/6$).  Yet the maximally spinning regime is of considerable
theoretical and observational interest.  Maximally spinning black
holes barely manage to hide their singularities within their event
horizons, and observable, ``naked'' singularities are expected for
spins $\chi > 1$.  While infinitesimal processes like steady accretion
cannot produce naked singularities \cite{Chandrasekhar:1985kt},
Penrose's cosmic-censorship conjecture \cite{Penrose:1969pc} that such
singularities can {\it never} be created has not been proven for
comparable-mass BBH mergers.  Black-hole spins are not a purely
theoretical concern; they can be measured by reverberation mapping of
iron K$\alpha$ fluorescence in the spectra of active galactic nuclei
(AGN) \cite{Reynolds:1998ie}.  This technique has been applied to {\it
  XMM-Newton} observations of the Seyfert 1.2 galaxy MCG-06-30-15,
leading to a measured spin $\chi = 0.989^{+0.009}_{-0.002}$ very near
the maximal limit \cite{Brenneman:2006hw}.  The spins of supermassive
black holes (SBHs) also offer important insights into their formation.
Some expect that highly spinning SBHs will only be found in gas-rich
systems like spirals \cite{Bogdanovic:2007hp}.  Others suggest that
gas accretion occurs through a series of chaotically oriented
episodes, leading to moderate spins ($\chi \sim 0.1 - 0.3$) lower than
those expected from comparable-mass mergers in gas-poor ellipticals
\cite{King:2008au}.  Comparing measured spins in spirals and
ellipticals will distinguish between these two scenarios.

In the absence of reliable simulations of maximally spinning BBH
mergers, we must rely on various approximations in this important
regime.  Hughes and Blandford (hereafter HB) \cite{Hughes:2002ei}
assumed that the energy and angular momentum radiated during the
inspiral stage dominates that radiated during the comparatively brief
merger and ringdown.  They then used conservation of energy and
angular momentum to equate these quantities when the BBHs reach their
innermost stable circular orbit (ISCO) to the mass and spin of the
final black hole,
\begin{subequations}
  \begin{eqnarray}  \label{E:E_HB}
      M_{f, {\rm HB}} &=& m_1 + m_2 E(\chi_1) \, , \\ \label{E:L_HB}
      S_{f, {\rm HB}} &=& m_1 m_2 L_{\rm orb}(\chi_1) + m_{1}^{2} \chi_1 \, .      
  \end{eqnarray}
\end{subequations}
Here $E(\chi)$ is the energy per unit mass of a test particle on an
equatorial orbit of a Kerr black hole with spin parameter $\chi$ and
$L_{\rm orb}(\chi)$ is the corresponding dimensionless orbital angular
momentum.  While $E(\chi)$ and $L_{\rm orb}(\chi)$ are only avaliable
in the test-particle limit ($m_2 \ll m_1$) \cite{Bardeen:1972fi}, HB
extended these exact, analytic results to comparable-mass mergers.
While this approach reproduces the test-particle limit by design, it
is not symmetric under the exchange of labels ``1'' and ``2'' and
leads to supermaximal spins ($\chi_f > 1$) for near maximal mergers.

Buonanno, Lehner, and Kidder \cite{Buonanno:2007sv} (hereafter BKL)
remedied these problems by (1) conserving mass, (2) including the
initial spin of the smaller black hole, and (3) using the spin
parameter of the {\it final} black hole $\chi_f$ rather than that of
the more massive of the initial BBHs $\chi_1$ to determine the orbital
angular momentum,
\begin{subequations}
  \begin{eqnarray} \label{E:E_BKL}
      M_{f, {\rm BKL}} &=& m_1 + m_2 \, , \\ \label{E:L_BKL}
      S_{f, {\rm BKL}} &=& m_1 m_2 L_{\rm orb}(\chi_f) + m_{1}^{2} \chi_1 +
      m_{2}^{2} \chi_2 \, .
  \end{eqnarray}
\end{subequations}
$L_{\rm orb}(\chi_f)$ in Eq.~(\ref{E:L_BKL}) corresponds to a prograde
(retrograde) equatorial orbit if the final spin is aligned
(anti-aligned).  This approach agrees remarkably well with existing
simulations of moderately spinning, comparable-mass BBH mergers, but
the assumption of mass conservation reduces agreement with the
test-particle limit to zeroth order in the mass ratio $q$.  While this
approximation is strictly valid to 10\% as noted by BKL, it
artificially reduces $\chi_f$ in the important highly spinning regime.

To improve on BKL, we seek an analytic expression for the final mass
$M_f$ for arbitrary mass ratios and initial spins.  In the spirit of
their work, we propose
\begin{equation} \label{E:E_K}
  M_{f, {\rm K}} = M - \mu [1 - E(\chi_f)] \, ,
\end{equation}
where $M \equiv m_1 + m_2$ is the total initial mass and $\mu \equiv
m_1 m_2/M$ is the reduced mass.  This proposal both agrees with
Eq.~(\ref{E:E_HB}) to first order in $q$ and is symmetric under
exchange of the black hole labels ``1'' and ``2''.  Dividing
Eq.~(\ref{E:L_BKL}) by the square of (\ref{E:E_K}), we find
\begin{equation} \label{E:chi_K}
  \chi_f = \frac{\nu L_{\rm orb}(\chi_f) + \frac{\chi_1}{4} (1 +
    \sqrt{1 - 4\nu})^2 + \frac{\chi_2}{4} (1 - \sqrt{1 - 4\nu})^2}{\{
    1 - \nu[1 - E(\chi_f)] \}^2} \, ,
\end{equation}
for the final spin parameter, where $\nu \equiv \mu /M$.

\begin{figure}[t!]
  \begin{center}
    \includegraphics[width=3.1in]{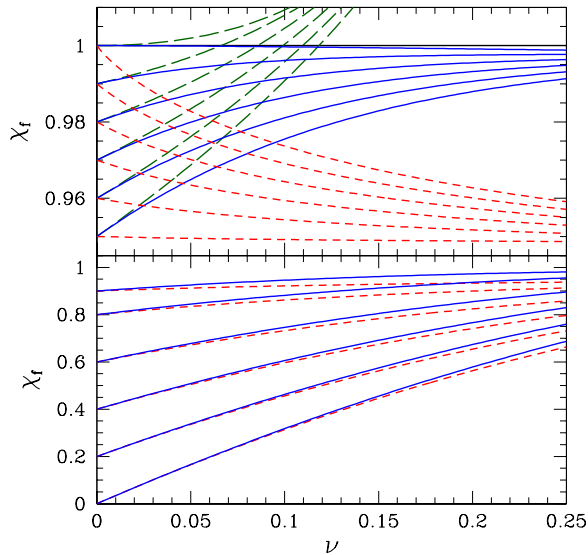}
  \end{center}
  \caption{The final spin parameter $\chi_f$ for mergers with equal
    initial spin parameters ($\chi_1 = \chi_2$) and mass ratio $\nu$.
    Each curve gives a different value of $\chi_1$, which can be read
    off the y-axis as $\chi_f = \chi_1$ for $\nu = 0$.  Curves in blue
    (solid) denote our predictions, red (short-dashed) those of BKL,
    and green (long-dashed) those of HB.  {\it Top panel}: the highly
    spinning regime ($\chi_i \geq 0.95$).  {\it Bottom panel}: aligned
    spins ($\chi_1 \geq 0$).}
  \label{F:EqChi}
\end{figure}

In Fig.~\ref{F:EqChi}, we compare this new prediction for $\chi_f$
(hereafter K) with those of HB (after including $S_2$ in
Eq.~(\ref{E:L_HB})) and BKL.  We consider BBHs with equal initial
spins ($\chi_1 = \chi_2 = \chi_i$) aligned with the orbital angular
momentum.  All three approaches reproduce the trivial zeroth order
result that $\chi_f = \chi_i$ in the test-particle limit ($\nu = 0$).
However, only K and HB provide the correct first-order behavior (the
value of $\partial\chi_f/\partial\nu$) in this limit \footnote{Correct
  dependence on $\nu$ helps calibrate fits for $\chi_f$; for example,
  Eq.~(\ref{E:chi_K}) implies the coefficients of
  \cite{Rezzolla:2007rd} should be $s_4 = -(2\sqrt{3})/9$, $t_0 =
  -(16\sqrt{3})/9$, and $t_1 = 2\sqrt{3}$.}.  The BKL assumption $M_f
= M$ implies that $E = 1$ and makes the denominator of
Eq.~(\ref{E:chi_K}) equal unity.  Overestimating $M_f$ to first order
in $\nu$ leads to a first-order underestimate in $\chi_f = S_f/M_f^2$,
accounting for the negative values of $\partial\chi_f/\partial\nu$
given by BKL for $\chi_i \gtrsim 0.95$.  K and HB predict
$\partial\chi_f/\partial\nu(\chi_i \leq 1) \geq 0$, with equality for
maximal initial spins.  This agrees with the famous result of Bardeen
\cite{Bardeen:1970} that black holes can be spun up to the Kerr limit
$\chi = 1$ after accreting a finite mass of test particles.  Despite
its success for $\nu = 0$, HB predicts naked singularities ($\chi_f >
1$) for comparable-mass mergers.

\begin{figure}[t!]
  \begin{center}
    \includegraphics[width=3.1in]{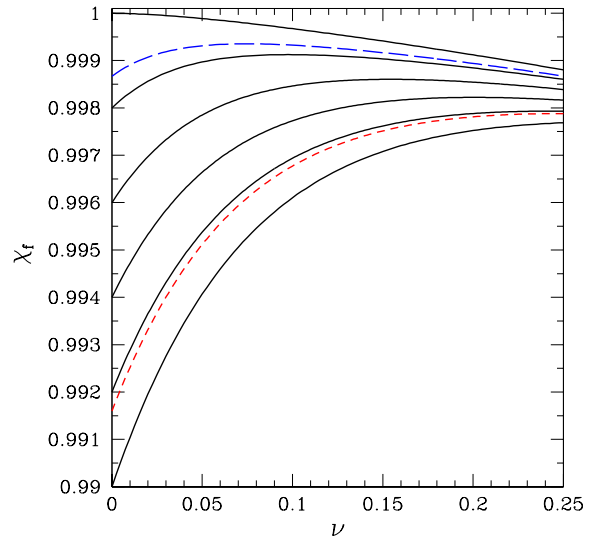}
  \end{center}
  \caption{The final spin parameter $\chi_f$ as a function of $\nu$
    for the merger of BBHs with equal and nearly maximal initial spin
    parameters ($\chi_1 = \chi_2 > 0.99$).  The red (short-dashed)
    curve shows the maximum initial spin ($\chi_i \simeq 0.9916$) for
    which this function monotonically increases with $\nu$.  The blue
    (long-dashed) curve shows the maximum initial spin ($\chi_i \simeq
    0.9987$) for which black holes are spun up by equal-mass mergers.}
  \label{F:ChiMaxNu}
\end{figure}

Let's examine the predictions of K more closely in the limit of nearly
maximal spins $(\chi_1 = \chi_2 > 0.99)$.  In Fig.~\ref{F:ChiMaxNu},
we show $\chi_f(\nu)$ at intervals of 0.002 in $\chi_i$.  For $\chi_i
\lesssim 0.9916$ (the red, short-dashed curve), $\chi_f(\nu)$ is
monotonically increasing implying that equal-mass mergers ($\nu =
0.25$) are the most efficient way of spinning up a black hole.  As
$\chi_i$ increases, the maximum $\nu_{\rm max}$ of $\chi_f(\nu)$ moves
to lower $\nu$. For $\chi_i \gtrsim 0.9987$ (the blue, long-dashed
curve) equal-mass mergers spin {\it down} black holes rather than
spinning them up.  Finally, at $\chi_i = 1$, the maximum reaches $\nu
= 0$ making $\chi_f(\nu)$ monotonically decreasing.  Thus maximally
spinning black holes will be spun down for all mergers with $\nu > 0$.
The merger of maximally spinning equal-mass BBHs yields a black hole
with $\chi_f \simeq 0.9988$.

\begin{figure}[t!]
  \begin{center}
    \includegraphics[width=3.1in]{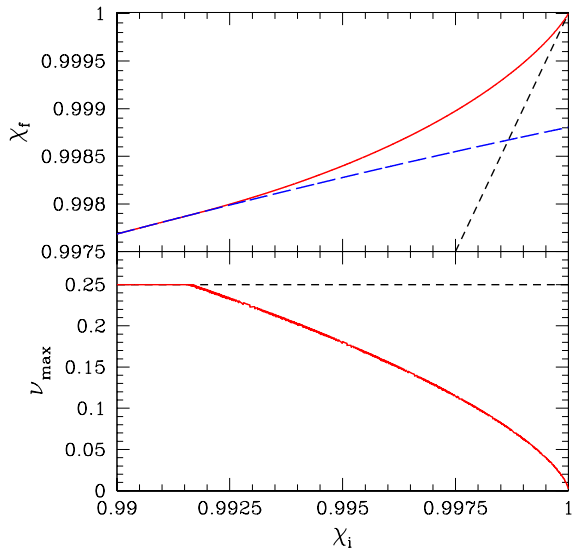}
  \end{center}
  \caption{{\it Top panel}: The final spin parameter $\chi_f$ as a
    function of $\chi_i$ for the merger of BBHs with equal and nearly
    maximal initial spin parameters ($\chi_1 = \chi_2 = \chi_i$).  The
    blue (long-dashed) curve corresponds to equal-mass BBHs, the red
    (solid) curve to mergers with mass ratio $\nu_{\rm max}$ that
    maximizes $\chi_f(\nu)$, and the black (short-dashed) curve to the
    reference $\chi_f = \chi_i$. {\it Bottom panel}: The value
    $\nu_{\rm max}(\chi_i)$ described above.}
  \label{F:ChiMaxChi}
\end{figure}

This behavior is summarized in Fig.~\ref{F:ChiMaxChi}.  At $\chi_i
\gtrsim 0.9916$, $\nu_{\rm max}$ falls below 0.25 and the red curve
peals away from the blue curve in the top panel, indicating that
equal-mass mergers are no longer the most efficient way of spinning up
highly spinning black holes.  While the blue curve belonging to
equal-mass mergers falls below the $\chi_f = \chi_i$ reference curve
for $\chi_i \gtrsim 0.9987$, the red curve always has $\chi_f \geq
\chi_i$ implying that BBH mergers can {\it always} spin up black holes
provided $\nu$ is low enough.  Though the specific values provided
here are only as accurate as the approximation itself, we believe this
qualitative behavior will ultimately be verified by numerical
simulations once near maximally spinning initial data can be
constructed.

\begin{figure}[t!]
  \begin{center}
    \includegraphics[width=3.1in]{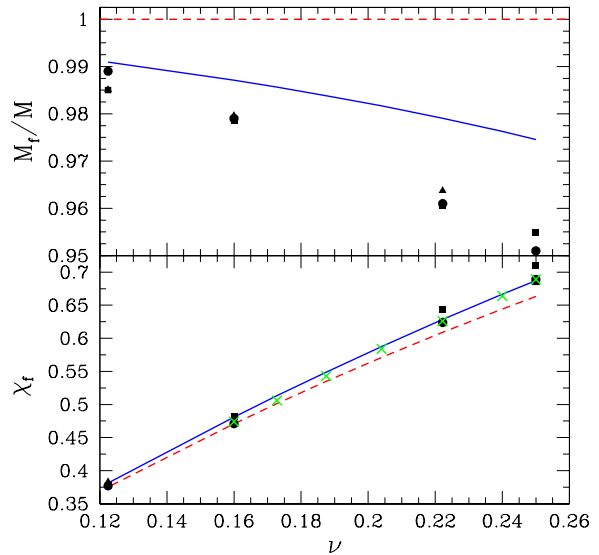}
  \end{center}
  \caption{{\it Top panel}: The final mass $M_f(\nu)$ for initally
    non-spinning mergers.  The red (short-dashed) curve is the BKL
    approximation, while the blue (solid) curve is that of
    Eq.~(\ref{E:E_K}).  The square, triangle, and circular points are
    three numerical estimates of $M_f$ \cite{Baker:2008mj}.  {\it
      Bottom panel}: The final spin $\chi_f(\nu)$ for non-spinning
    mergers.  The red (short-dashed) curve is the BKL prediction; the
    blue (solid) curve is that of Eq.~(\ref{E:chi_K}).  The black
    points are three estimates of the final spins from
    \cite{Baker:2008mj}, while the green Xs are simulated final spins
    from \cite{Berti:2007fi}.}
  \label{F:NonS}
\end{figure}

Let's compare these predictions to recent numerical-relativity
simulations.  We begin with initially non-spinning BBHs of mass ratio
$\nu$.  The top panel of Fig.~\ref{F:NonS} shows the final masses
$M_f$ determined from the energy radiated (squares), apparent horizon
(triangles), and quasi-normal modes (QNMs, circles)
\cite{Baker:2008mj}.  Eq.~(\ref{E:E_K}) underestimates $M_f$ by nearly
a factor of two for equal-mass mergers, possibly indicating that as
much as 50\% of the energy is radiated after ISCO for these mergers.
Better agreement is seen at lower mass ratios, in line with arguments
that the energy radiated during merger and ringdown scales as $\nu^2$
rather than $\nu$ \cite{Baker:2008mj,Berti:2007fi}.  Despite these
poor estimates of $M_f$, our predicted final spins $\chi_f$ are
amazingly accurate; the equal-mass prediction ($\chi_f = 0.687$) lies
within the numerical errors ($\Delta\chi_f \simeq 0.002$) of most
simulations, an order-of-magnitude improvement over BKL.  This is
telling us something profound; while significant energy and angular
momentum are radiated after plunge, they are radiated in the
appropriate ratio to preserve the spin $\chi_f = S_f/M_f^2$.

\begin{figure}[t!]
  \begin{center}
    \includegraphics[width=3.1in]{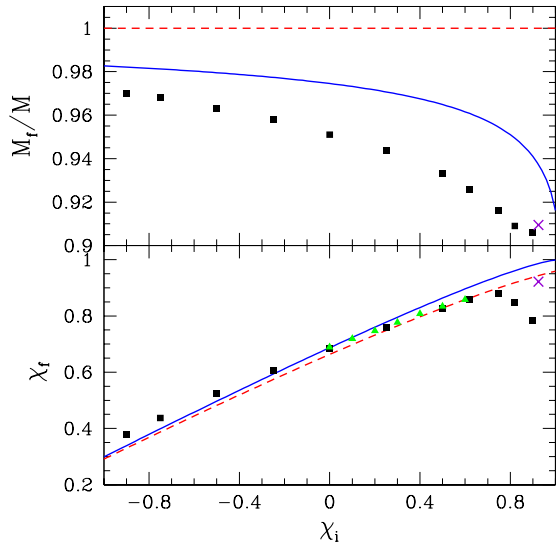}
  \end{center}
  \caption{{\it Top panel}: The final mass $M_f(\chi_i)$ for
    equal-mass, equal-spin mergers.  The red (short-dashed) curve
    is the BKL prediction, while the blue (solid) curve is
    that of Eq.~(\ref{E:E_K}).  Square points are simulations from
    \cite{Marronetti:2007wz}; the purple X is from \cite{Dain:2008ck}.
    {\it Bottom panel}: The final spin parameter $\chi_f(\nu)$ for these
    mergers.  The red (short-dashed) curve is the BKL prediction, the
    blue (solid) curve is that of Eq.~(\ref{E:chi_K}).  Black squares,
    purple Xs, and green triangles are from \cite{Marronetti:2007wz},
    \cite{Dain:2008ck}, and \cite{Rezzolla:2007xa}.}
  \label{F:EqSim}
\end{figure}

This is not always the case for spinning BBHs.  Fig.~\ref{F:EqSim}
shows the final masses and spins for equal-mass, equal-spin BBH
mergers.  If we believe our approximation, as much as 45\% of the
energy is radiated after the inspiral stage for aligned, highly
spinning configurations.  This energy loss tends to drive $\chi_f$ up,
but apparently angular momentum is radiated even more efficiently
after plunge because the numerically determined spins are below our
predictions.  The BKL approximation actually does better for these
configurations, but this appears to be a conspiracy between the excess
final mass compensating for the failure to account for the angular
momentum radiated after plunge.  Angular momentum is radiated less
efficiently after plunge for anti-aligned spins, leading us to
underestimate $\chi_f$ for $\chi_i < 0$.  Future work will attempt to
improve our approximation by accounting for radiation after the ISCO.

\begin{figure}
  \begin{center}
    \includegraphics[width=3.1in]{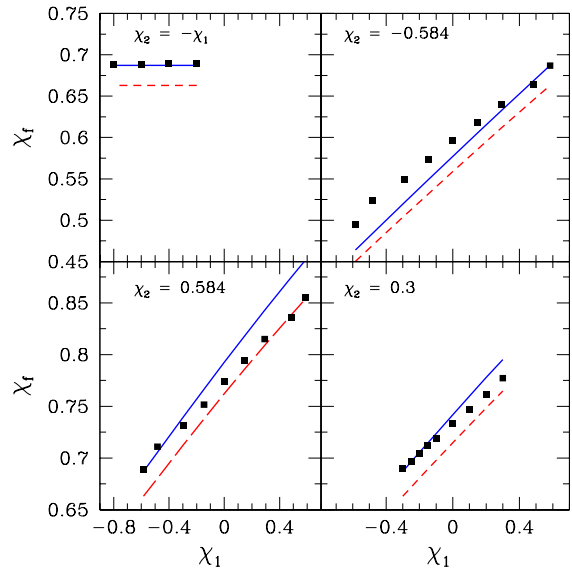}
  \end{center}
  \caption{Final spin $\chi_f$ as a function of $\chi_1$ for the four
    sequences of equal-mass, unequal-spin simulations presented in
    \cite{Rezzolla:2007xa}.  The value of $\chi_2$ for each sequence
    is listed on each panel, and we have exchanged $\chi_1$ and
    $\chi_2$ in the upper right panel for convenient presentation.
    Blue (solid) curves are our predictions, red (dashed) curves those
    of BKL.}
  \label{F:RezzSim}
\end{figure}

This same trend is seen in equal-mass, unequal-spin simulations
\cite{Rezzolla:2007xa}.  There is excellent agreement between our
approach and all simulations with vanishing {\it total} initial spin
like those shown in the upper left panel of Fig.~\ref{F:RezzSim}.  Our
approach improves on BKL except for the highly spinning aligned
configurations where their artificially high $M_f$ drives $\chi_f$
down.

BKL used conservation of angular momentum to predict with great
success the final spin in BBH mergers.  We have improved on their
approach by allowing for energy loss during the inspiral, and have
uncovered qualitatively new dependence of the final spin on the spins
and mass ratio of the initial BBHs.  Equal-mass BBH mergers might lead
to higher final spins than expected, rivaling the highest spins ($\chi
\simeq 0.998$) allowed by physical accretion flows
\cite{Thorne:1974ve}.  Black holes can {\it always} be spun up by
aligned mergers provided the mass ratio is small enough.  We have
found that for non-spinning BBH mergers, exact analytic results for
test particles on Kerr geodesics can be extrapolated all the way to
equal masses with better than 1\% accuracy.  Our approach provides
quantitative predictions for regimes not yet accessible to numerical
relativity, and helps guide the choice of simulations to explore the
7-dimensional parameter space of BBH mergers formed by $\nu$,
$\vec{S}_1$, and $\vec{S}_2$.  This complements fitting formulae that
successfully describe final spins for comparable-mass mergers
\cite{Rezzolla:2007xa,Rezzolla:2007rd,Rezzolla:2007rz}.  It can help
identify where in parameter space black holes will be spun up by
mergers, and where the final spin will vanish leading to interesting
orbital dynamics.  Understanding BBH mergers is an essential goal of
astrophysics, general relativity, and gravitational-wave detection,
and can only be achieved by combining numerical relativity,
post-Newtonian techniques, and analytic approximations like ours.


{\bf Acknowledgements.}  We thank Emanuele Berti, Latham Boyle, Avery
Broderick, Alessandra Buonanno, Larry Kidder, Luis Lehner, Harald
Pfeiffer, Denis Pollney, and Luciano Rezzolla for useful
conversations.

\end{document}